# Mechanical photoluminescence excitation spectra of a strongly driven spin-mechanical system

Xinzhu Li and Hailin Wang

Department of Physics, University of Oregon, Eugene, OR 97403, USA

## Abstract

We report experimental studies of a driven spin-mechanical system, in which a nitrogen vacancy (NV) center couples to out-of-plane vibrations of a diamond cantilever through the excited-state deformation potential. Photoluminescence excitation studies show that in the unresolved sideband regime and under strong resonant mechanical driving, the excitation spectra of a NV optical transition feature two spectrally sharp peaks, corresponding to the two turning points of the oscillating cantilever. In the limit that the strain-induced frequency separation between the two peaks far exceeds the NV zero-phonon linewidth, the spectral position of the individual peak becomes sensitive to minute detuning between the mechanical resonance and the external driving force. For a fixed optical excitation frequency near the NV transition, NV fluorescence as a function of mechanical detuning features resonances with a linewidth that can be orders of magnitude smaller than the intrinsic linewidth of the mechanical mode. This enhanced sensitivity to mechanical detuning can potentially provide an effective mechanism for mechanical sensing, for example, mass sensing via measurements of induced changes in the mechanical oscillator frequency.

# I. INTRODUCTION

In a driven spin-mechanical system, an electron spin or more generally an artificial atom, such as a defect center, couples to driven vibrations of a mechanical oscillator via direct or phonon-assisted transitions[1, 2]. Driven spin-mechanical systems have been exploited for applications such as the mechanical quantum control of electron or nuclear spins and the suppression of spin dephasing via dressed spin states[3-14]. For spin-mechanical systems that feature phonon-assisted (i.e., sideband) transitions, earlier experimental studies have emphasized the resolved-sideband regime, for which the mechanical frequency is large compared with the transition linewidth, since this regime is desirable for cooling and amplification of mechanical motion via spin-mechanical coupling[15, 16], as well as for applications in quantum control, though experimental studies in the unresolved sideband regime have been carried out in NV-based as well as quantum dot-based spin-mechanical systems[17-19].

In this paper, we report experimental studies of a driven spin-mechanical system in the unresolved sideband regime. For our system, a nitrogen vacancy (NV) center in a diamond cantilever couples to out-of-plane modes of the cantilever via the deformation potential of the NV excited states and through the sideband optical transitions. Photoluminescence excitation (PLE) studies of the NV center show that under a strong external driving of the mechanical mode, the excitation spectra from a NV optical transition feature two spectrally sharp peaks with a large strain-induced frequency separation, in contrast to the multiple sidebands observed in earlier experimental studies in the resolved sideband regime[20, 21]. While the frequency separation increases linearly with the amplitude of the mechanical vibration, the linewidth of the peaks remains close to that of the zero-phonon line (ZPL). In the limit that the strain-induced frequency separation far exceeds the ZPL linewidth, NV fluorescence at a given optical excitation frequency near the NV transition becomes sensitive to minute detuning between the mechanical resonance and external driving force, even when the detuning is far smaller than the intrinsic linewidth of the mechanical mode. The resulting mechanical PLE spectra, i.e., NV fluorescence as a function of mechanical detuning, feature resonances with a linewidth that can be orders of magnitude smaller than the intrinsic linewidth of the mechanical mode.

The greatly enhanced sensitivity to mechanical detuning can provide an effective mechanism for mechanical sensing [22], specifically, by monitoring frequency shifts of the mechanical oscillator through the sharp resonances in mechanical PLE. For earlier mechanical sensing studies, mechanical frequencies are monitored via piezo-resistive, capacitive, optical interferometric, or optomechanical measurements. These measurements are limited by the intrinsic linewidth of the relevant mechanical mode.

## II. EXPERIMENTAL METHOD

Diamond cantilevers used in this study feature a width of 4 μm, length near 15 μm, and a thickness near 2.5 μm. The cantilevers are embedded in a square phononic crystal lattice, as shown in Fig. 1a. A phononic band gap of the square lattice protects the fundamental out-of-plane mode of the cantilevers from its surrounding environment[23]. The diamond phononic structure is fabricated from electronic grade single-crystal bulk diamond with {100} faces (from Element Six, Inc.), with electron-beam lithography and reactive ion etching. NV centers are created about 100 nm below the diamond surface with ion implantation followed by stepwise high temperature thermal annealing. Details of the sample fabrication are presented in earlier studies[24, 25].

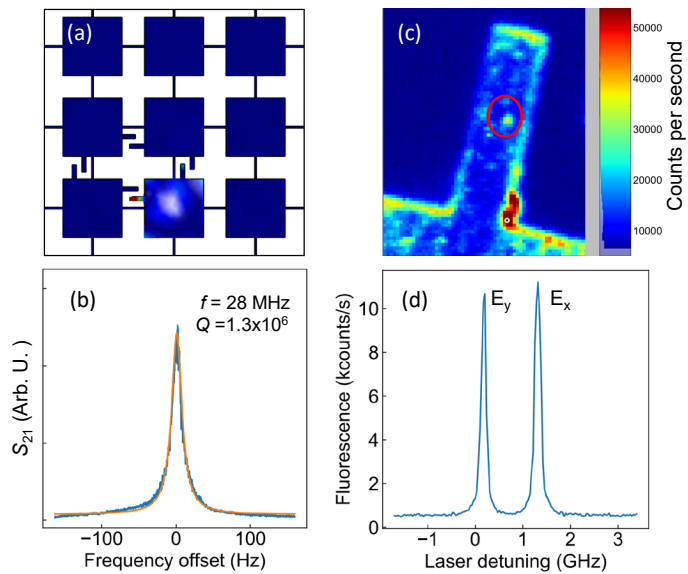

FIG. 1. (a) Simulated displacement pattern of diamond cantilevers embedded in a square phononic crystal lattice with a period of 76 μm. The width and length of the bridge is 1.3 and 20 μm, respectively. (b) Spectral response of the fundamental out-of-plane mode of the cantilever used. (c) Confocal optical image of the cantilever. The red circle highlights the NV center used in the PLE study. (d) PLE spectrum of the NV center. The two ZPLs correspond to optical transitions from the $m_s=0$ ground state to the $E_x$ and $E_y$ excited states.

For the experimental results presented in this paper, the fundamental out-of-plane mode of the cantilever has a frequency, $\omega_m/2\pi$=28 MHz, and a linewidth, $\gamma_m/2\pi$=22 Hz, corresponding to a Q-factor, $Q$=1.3x10$^6$ (see Fig. 1b). The high Q-factor results from the protection of the phononic band gap. The characterization of the mechanical modes is discussed in detail in our earlier study [23]. PLE studies were carried out in a NV center near the middle of the cantilever, as shown in the confocal optical image in Fig. 1c. The PLE spectrum of the NV center obtained at T=8 K and in the absence of external mechanical driving is characterized by two ZPLs, corresponding to the transitions from the $m_s$=0 ground state to the $E_x$ and $E_y$ excited states (see Fig. 1d)[26]. The ZPLs feature a spectral linewidth near 100 MHz, indicating the high optical quality of the NV center, though the spin-mechanical coupling via the sideband optical transitions is in the unresolved sideband regime.

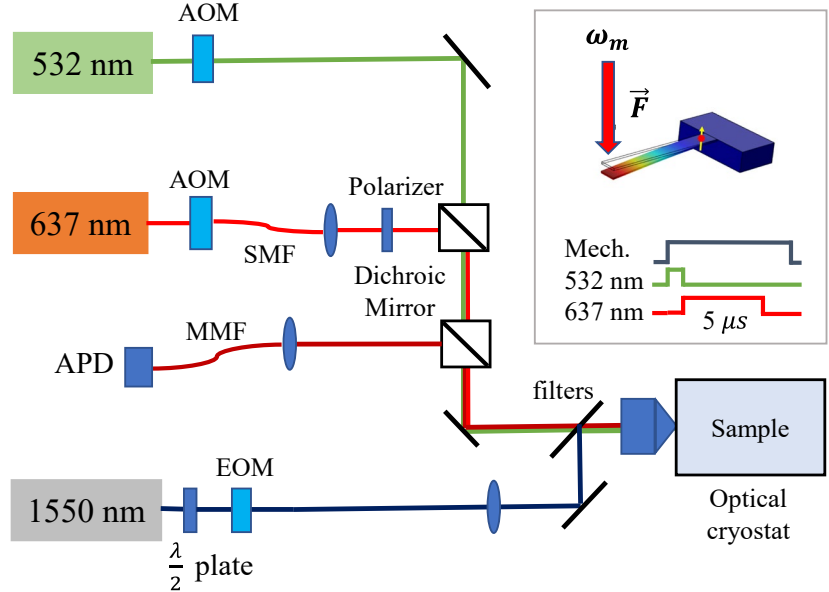

FIG. 2. Schematic of the experimental setup. AOM: Acousto-optic modulator. APD: Avalanche photodiode. MMF: Multimode fiber. SMF: Single mode fiber. The inset illustrates a schematic of resonantly driving the vibrations of the cantilever with radiation pressure force and shows the pulse sequence for the PLE experiment.

Figure 2 shows a schematic of the experimental setup, which is based on a home-built low-temperature scanning confocal microscope. The thin diamond phononic structure is glued to a bulk diamond plate mounted on the cold finger of an optical cryostat (Montana Instrument S50). All experimental results presented in this paper were obtained with an estimated sample temperature of 8 K unless otherwise specified. For the PLE experiment, a green laser pulse ($\lambda$=532 nm) initializes the NV center into the $m_s$=0 ground state. A tunable diode laser (New Focus Velocity TLB-6712) near 637.2 nm excites the NV center. To avoid the power broadening of the ZPL, the red laser power is kept below 1 µW. NV fluorescence with $\lambda$ > 645 nm is collected with a 100x objective (Nikon L Plan 0.85 NA) and then coupled into an optical fiber connecting to an avalanche photodiode (APD) for single-photon counting. A pulse sequence of the PLE experiment is shown in the inset of Fig. 2. Photon counting takes place when the 637.2 nm laser is on. An averaging time of 0.5 s is used for each data point in a PLE spectrum.

To resonantly drive the mechanical modes of the cantilever, we use radiation pressure force from a focused laser beam with $\lambda$=1.55 µm, for which the intensity of the laser beam is sinusoidally modulated with an electro-optical modulator (EOM) and with a 100% modulation depth. An additional lens is also used such that the 1.55 µm laser beam can be focused onto the cantilever surface along with the green and red laser beams. Within the range of the laser power used (<50 mW), no ZPL broadening due to sample heating induced by the 1.55 µm laser (with the intensity modulation off) has been observed. Note that optical powers given in the paper were measured before the 100x objective in the setup.

## III. EXPERIMENTAL RESULTS AND ANALYSIS

### A. Photoluminescence excitation of a driven spin-mechanical system

To investigate the coupling between the NV center and the driven mechanical vibration, we measure the PLE spectra of the NV center, while resonantly driving the fundamental out-of-plane mode. As shown in Fig. 3a, at relatively weak mechanical driving, a broadening of the $E_y$ resonance is observed, which is expected since the spin-mechanical system is in the unresolved band regime. With increasing mechanical driving and with other conditions remaining unchanged, two peaks appear in the PLE spectra of the $E_y$ transition (see Fig. 3b). Similar behaviors have also been observed in an earlier study[17]. The frequency separation between the two peaks is linearly proportional to the amplitude of the driving force, in this case, the average power of the 1.55 µm laser beam (see Fig. 3c). Overall, the excitation spectra of the spin-mechanical system in the unresolved sideband regime differ qualitatively from those observed in the resolved sideband regime, which are characterized by distinct sideband resonances corresponding to emission or absorption of integer numbers of phonons [21].

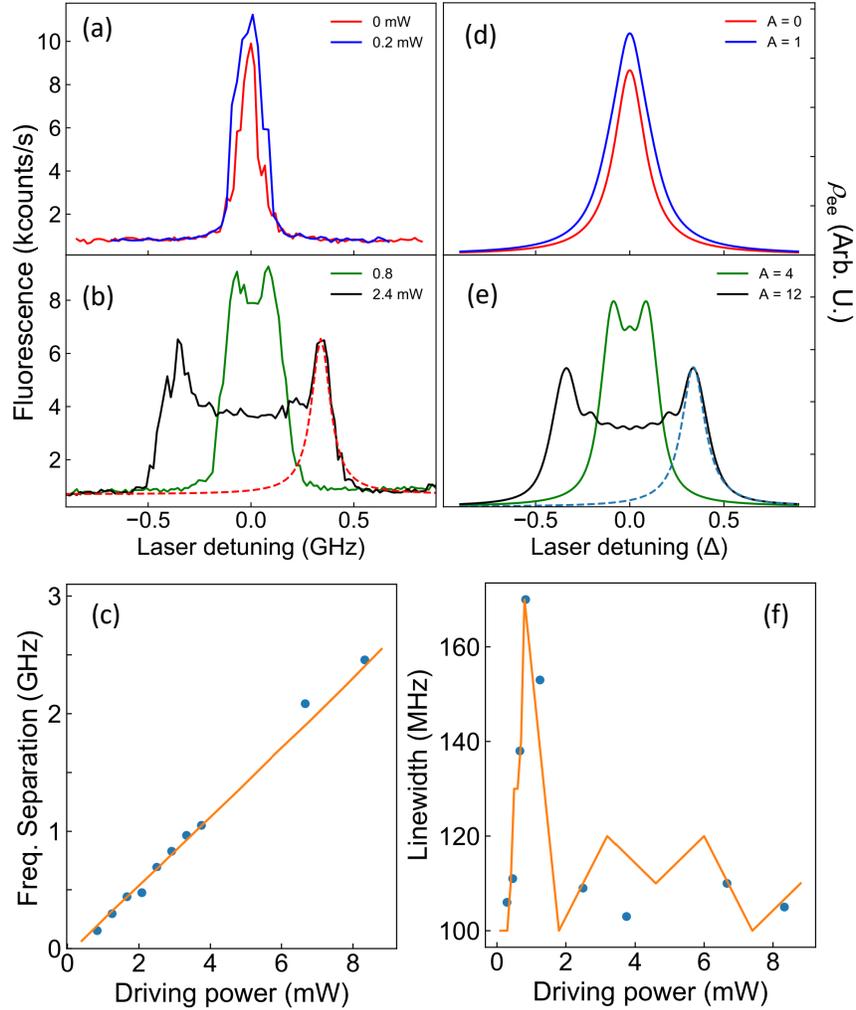

Fig 3. (a) and (b) PLE spectra of the NV center under resonant mechanical driving, with the 1.55 μm laser average power indicated in the figure. (c) Frequency separations of the two peaks in PLE spectra as a function of the mechanical driving power. (d) and (e) Theoretically calculated PLE spectra as discussed in the text. The dashed line is a Lorentzian fit to the steep side of the peak. (f) Spectral linewidth of the PLE peaks versus the mechanical driving power. The solid lines in (c) and (f) are derived from the theoretically calculated PLE spectra, for which a conversion factor of 0.2 mW laser power corresponding to $A$=1 is used.

To understand the excitation spectrum of the driven mechanical system in the unresolved sideband regime, we model the NV center as a two-level system, which couples to the long wavelength mechanical vibration through the excited-state deformation potential, $D$. The electron-phonon interaction Hamiltonian describing the strain-induced energy shift of the excited state, $|e>$, can be written as [20, 27]

$$V_{e-phonon} = \hbar A \omega_m \sin(\omega_m t + \phi) |e><e|, \qquad (1)$$

where $A$ is a dimensionless driving amplitude proportional to both $D$ and the amplitude of the mechanical vibration. For resonant or nearly resonant optical excitations, this interaction Hamiltonian leads to sideband optical transitions corresponding to the absorption or emission of integer numbers of phonons. To calculate the excited state population of the NV center, we solved the density matrix equations in the steady state, assuming that the NV is initially in the $m_s=0$ ground state and $\omega_m$ is large compared with the excited-state population decay rate. In the limit of weak optical excitation, the excited state population for an optical excitation field with frequency $\omega$ is given by

$$\rho_{ee}(\omega) \propto \sum_{n=-\infty}^{+\infty} \frac{J_n^2(A)}{(\omega - \omega_0 - n\omega_m)^2 + (\gamma/2)^2}, \qquad (2)$$

where $\gamma$ and $\omega_0$ are the linewidth and resonance frequency of the ZPL, respectively, and $J_n(x)$ is the Bessel function of the first kind. As shown in Eq. 2, the excitation of the NV center includes the contributions from all relevant sidebands, i.e., phonon-assisted optical transitions, with a relative weighting determined by $J_n^2(A)$. Figures 3d and 3e show the theoretically calculated PLE spectra obtained for $A=0, 1$ and for $A=4, 12$, respectively, with $\gamma/2\pi= 100$ MHz. For $A>>1$, the frequency separation between the two PLE peaks derived from the calculation is linearly proportional to $A$. The small oscillations in Fig. 2e are remnants of phonon sidebands. For a direct comparison between the experimental and theoretical results, we plot in Fig. 3c the calculated frequency separation with a fixed conversion factor between $A$ and the power of the 1.55 μm laser. A good agreement between the theory and experiment is achieved with 0.2 mW laser power corresponding to $A=1$. The same conversion factor is also used for the theoretical results shown in Fig. 3f.

The linear scaling between the frequency separation of the two peaks in the PLE spectra and the amplitude of the external driving force shown in Fig. 3c suggests a classical interpretation of the PLE spectra at relatively large $A$. In this case, the frequency separation corresponds to the strain-induced relative shift of the excited state between the two turning points, i.e., the maximum displacements of the cantilever vibration. Furthermore, the excitation spectra at the turning points are expected to retain the spectrally sharp NV zero-phonon resonance except for a spectral shift. This is confirmed by the experimental and theoretical results shown respectively in Figs. 3b and 3e. The experimental results on the linewidth of the PLE peaks as a function of the 1.55 μm laser power shown in Fig. 3f provide additional confirmation. As shown in Fig. 3f, after a large increase in the linewidth, which occurs before the two peaks appear, the linewidth at relatively large $A$ returns to values, which only slightly exceed the ZPL linewidth. The measured linewidths are also in general agreement with the calculated linewidths plotted in Fig. 3f. For the asymmetric PLE peaks, we have used a Lorentzian fit from the steep side of the peak to extract the linewidth (see Figs. 3b and 3e). The variations in the linewidths under strong mechanical driving shown in Fig. 3f are to a large extent due to the variations in the curve fitting. While other methods can also be used to extract the linewidth, the exact linewidth, other than whether it is close to the ZPL linewidth, does not provide much additional information.

It should be noted that results similar to those shown in Fig. 3 have also been observed for the $E_x$ transition, for which the strain-induced frequency separation between the two peaks in PLE spectra is about 50% of that observed for the $E_y$ transition. In the limit that the frequency separation far exceeds the ZPL linewidth, nearly the same linewidth of the PLE peaks have been observed for both transitions.

**B. Mechanical photoluminescence excitation**

In the limit that the strain-induced frequency separation between the two PLE peaks far exceeds the ZPL linewidth, the PLE spectra depends sensitively on mechanical detuning (i.e., the detuning between the modulation frequency of the 1.55 μm laser and the resonance frequency of the mechanical mode), even when the detuning is small compared with $\gamma_m$, as shown in Figs. 4a and 4b. Figures 4c and 4d plot the NV fluorescence obtained as a function of the mechanical detuning at two different 1.55 μm laser powers, where the optical excitation frequency is fixed near a peak of the corresponding PLE spectrum obtained at zero mechanical detuning.

The NV fluorescence vs mechanical detuning shown in Figs. 4c and 4d can be viewed as mechanical PLE (mPLE), for which varying the mechanical detuning effectively shifts the resonance frequency of the NV optical transition. These spectra are characterized by two sharp resonances. Each resonance occurs at a mechanical detuning, for which the spectrally shifted optical transition at a turning point of the mechanical oscillation is resonant with the optical excitation field. The mPLE spectra shown in Figs. 4c and 4d feature the unusual behavior that the linewidth of the individual resonance, $\gamma_{mPLE}$, can be orders of magnitude smaller than the intrinsic mechanical linewidth, $\gamma_m$. Furthermore, the stronger the external mechanical drive, the sharper the resonance becomes.

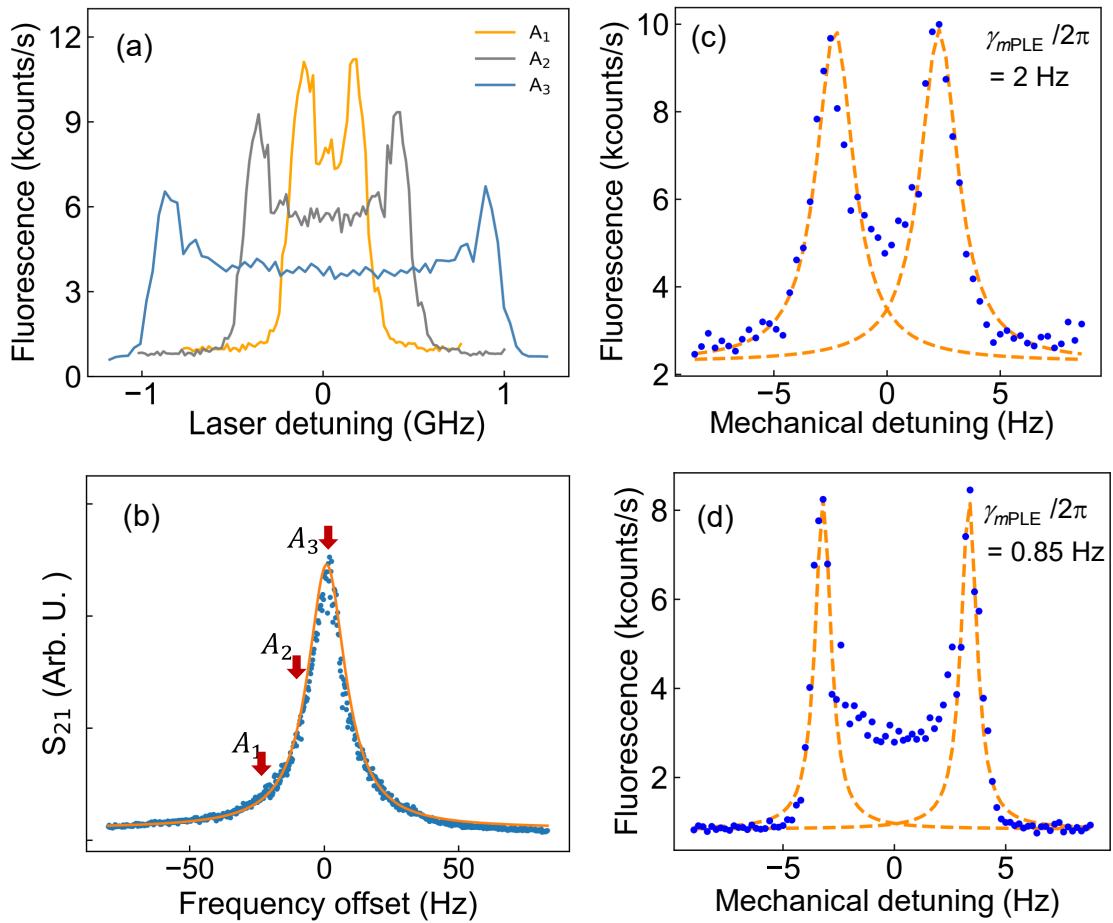

Fig 4. (a) PLE spectra of the NV center obtained with a 1.55 μm laser power of 6 mW, with the mechanical detuning indicated in (b). (b) Mechanical frequencies (indicated by the arrows) used for the PLE spectra in (a) relative to the resonance of the fundamental out-of-plane mode of the cantilever. (c) NV fluorescence as a function of the frequency

detuning of the mechanical drive, obtained with a 1.55 μm laser power of 9 mW and with the optical excitation frequency fixed near a peak of the corresponding PLE spectrum obtained at zero mechanical detuning. The dashed lines are Lorentzian fits to the steep side of the resonances. (d) The same as (c) except for a 1.55 μm laser power of 14 mW.

As shown in Fig. 4a, the spectral position, $s$, of the NV optical transition at a cantilever turning point shifts with mechanical tuning, $\delta$. For a mPLE resonance centered at $\delta_0$, the linewidth of the resonance is determined by the mechanical detuning needed to induce a frequency shift of $\gamma$ for the corresponding NV transition, i.e.,

$$\gamma_{mPLE} \cdot \left.\frac{ds}{d\delta}\right|_{\delta=\delta_0} \approx \pm\gamma. \tag{3}$$

In the limit that $S/\gamma \gg 1$, where $S$ is the strain-induced frequency separation between the two peaks in the PLE spectra obtained at $\delta=0$, we have

$$\gamma_{mPLE} \approx \frac{(4\delta_0^2 + \gamma_m^2)^2}{4|\delta_0|\gamma_m^2} \cdot \frac{\gamma}{S} \tag{4}$$

For Figs. 4c and 4d, $|\delta_0|/2\pi$=2.3 and 3.5 Hz and $S/2\pi$=2.97 and 4.62 GHz (derived from the linear dependence between $S$ and the 1.55 μm laser power in Fig. 3c), respectively. Theoretical estimates using Eq. 4 yield $\gamma_{mPLE}/\gamma_m = 0.088$ and 0.041, in agreement with the experimentally obtained linewidth reduction, $\gamma_{mPLE}/\gamma_m = 0.091$ and 0.039, for Figs. 4c and 4d, respectively.

The sharp resonance in a mPLE spectrum can provide an effective mechanism to enhance the sensitivity of mechanical oscillator-based sensing[22], for example, mass sensing by monitoring the frequency shift of the mechanical oscillator[28-30]. The diamond cantilever used in this study has a mass near 0.5 nanogram. A mass change of 1 attogram results in a frequency shift of order 0.03 Hz. As shown in Fig. 4d, the linewidth of the resonances in the mPLE is 0.85 Hz, compared with the intrinsic mechanical linewidth of 22 Hz. A greater reduction in the linewidth can be achieved with a stronger external mechanical drive. For the experimental implementation of mechanical sensing, a straightforward approach is to monitor the NV fluorescence while fixing the frequency of the nearly resonant external mechanical drive near the middle of the steep slope of a resonance in the mPLE spectrum.

To determine the sensitivity for measuring mechanical frequency shifts, we monitor the NV fluorescence while sinusoidally varying the mechanical detuning (see Fig. 5), where the zero frequency-detuning corresponds to the midpoint of the steep side of a mPLE resonance. The sensitivity for the frequency-shift measurement can be defined as [31]

$$\eta = \frac{\sigma^{1s}}{|dF/d\delta|} \tag{5}$$

where $\sigma^{1s}$ is the uncertainly for the fringe contrast $F$ for 1 second averaging and $dF/d\delta$ is the gradient of $F$ with respect to the mechanical detuning. From the results shown in Fig. 5, we estimate that $\eta = 0.01$ and $0.02$ $Hz/\sqrt{Hz}$ when a 1.55 μm laser power of 11 and 6 mW is used, respectively. The better sensitivity at stronger mechanical driving arises from the increase in the gradient due to the reduction in $\gamma_{mPLE}$. For comparison, frequency noise floor for traditional mass sensing experiments using cantilevers with $Q < 5000$ typically exceeds $0.1$ $Hz/\sqrt{Hz}$ [28, 29]. Note that the sensitivity shown in Fig. 5 is limited by the relatively small photon count rate used in the experiment and can thus be significantly improved.

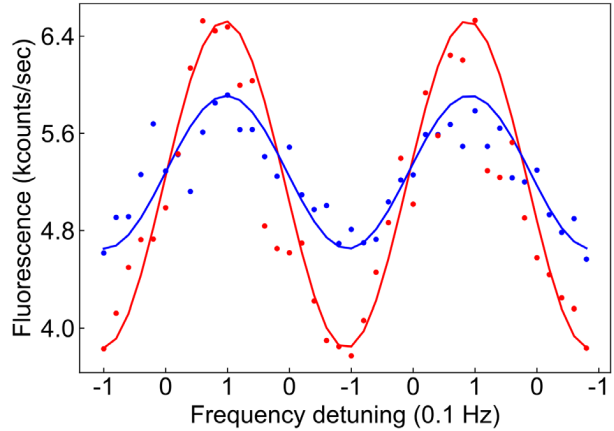

Fig 5. NV Fluorescence as a function of oscillating mechanical detuning. The solid lines are least-square fits to sinusoidal oscillations. For red (and blue) dots, a 1.55 μm laser power of 11 (and 6) mW is used, with other conditions nearly the same as those for Fig. 4.

**IV. CONCLUSION**

In summary, using a diamond cantilever as a model spin-mechanical system, we have shown that in the unresolved sideband regime and under strong resonant mechanical driving, the excitation spectra of a NV optical transition feature two sharp peaks, corresponding to the two turning points of the oscillating cantilever. In the limit that the strain-induced frequency separation between the two peaks is large compared with the ZPL linewidth, the spectral position of the individual peak becomes highly sensitive to minute detuning of the mechanical drive, leading to sharp resonances in mechanical PLE spectra with a linewidth, which can be orders of magnitude smaller than the intrinsic mechanical linewidth. This enhanced sensitivity to mechanical detuning can provide an effective mechanism for mechanical sensing, for example, mass sensing via measurements of induced mechanical frequency shifts. While a NV center has been used in this study, similar spin-mechanical processes and their potential applications can also be pursued in other defect or material systems.

**ACKNOWLEDGEMENTS**

This work is supported by the National Science Foundation (NSF) under Grant No. 2012524. We thank Kai-Mei Fu and Srivatsa Chakravarthi for the use of the diamond high-temperature annealing facility at the University of Washington.